\documentclass[structabstract]{aa}
\usepackage{epsfig}

\usepackage{graphicx}
\usepackage{longtable}
\usepackage{txfonts}

\begin{document}

   \title{High energy variability of 3C 273 during the AGILE multiwavelength campaign of December 2007 - January 2008}

   \subtitle{}

   \author{L. Pacciani \inst{1}
     \and
          I. Donnarumma \inst{1}
     \and
          V. Vittorini \inst{1}   \inst{- 4}
     \and
          F. D'Ammando \inst{1}   \inst{- 5}
     \and  \\
          M. T. Fiocchi \inst{1}
      \and
          D. Impiombato \inst{9}
     \and
          G. Stratta \inst{3}
     \and
     F. Verrecchia \inst{3}
     \and \\
          A. Bulgarelli \inst{6}
     \and
          A. W. Chen \inst{2} \inst{- 4}
     \and
          A. Giuliani \inst{2}
     \and
          F. Longo \inst{7}
     \and
          G. Pucella \inst{1}
     \and
          S. Vercellone \inst{2}
     \and \\
          M. Tavani \inst{1}
     \and
          A. Argan \inst{1}
     \and
          G. Barbiellini \inst{7}
     \and
          F. Boffelli \inst{8}
     \and
          P. A. Caraveo \inst{2}
     \and
          P. W. Cattaneo \inst{8}
     \and \\
          V. Cocco \inst{4}
     \and
          E. Costa \inst{1}
     \and
          E. Del Monte \inst{1}
     \and
          G. Di Cocco \inst{6}
     \and
          Y. Evangelista \inst{1}
     \and \\
          M. Feroci \inst{1}
     \and
          T. Froysland \inst{4}
     \and
          F. Fuschino \inst{6}
     \and
          M. Galli \inst{10}
     \and
          F. Gianotti \inst{6}
     \and \\
          C. Labanti \inst{6}
     \and
          I. Lapshov \inst{1}
     \and
          F. Lazzarotto \inst{1}
     \and
          P. Lipari \inst{11}\
     \and
          M. Marisaldi \inst{6}
     \and
          S. Mereghetti \inst{2}
     \and \\
          A. Morselli \inst{12}
     \and
          A. Pellizzoni \inst{2}
     \and
          F. Perotti \inst{2}
     \and
          P. Picozza \inst{12}
     \and
          M. Prest \inst{13}
     \and
          M. Rapisarda \inst{14}
     \and \\
          P. Soffitta \inst{1}
     \and
          M. Trifoglio \inst{6}
     \and
          G. Tosti \inst{9}
     \and
          A. Trois \inst{1}
     \and
          E. Vallazza \inst{7}
     \and \\
          D. Zanello \inst{11}
     \and
          L. A. Antonelli \inst{3}
     \and
          S. Colafrancesco \inst{3}
     \and
          S. Cutini \inst{3}
     \and \\
          D. Gasparrini \inst{3}
     \and
          P. Giommi \inst{3}
     \and
          C. Pittori \inst{3}
     \and
     L. Salotti \inst{15}
        }
   \institute{
              INAF/Istituto di Astrofisica Spaziale e Fisica Cosmica di Roma,
              Via Fosso del Cavaliere, 100 - I-00133 Roma, Italy
              \\\email{luigi.pacciani@iasf-roma.inaf.it}
              \and
              INAF/Istituto di Astrofisica Spaziale e Fisica Cosmica di Milano,
              Via E. Bassini, 15 - I-20133 Milano, Italy
              \and
              ASI Science Data Center,
              Via G. Galilei, I-00044 Frascati (Roma), Italy
              \and
              Consorzio Inter-universitario Fisica Spaziale,
              Viale Settimio Severo 3, I-10133, Torino, Italy
              \and
              Dip. di Fisica, Univ. Tor Vergata,
              Via della Ricerca Scientifica 1, I-00133 Roma, Italy,
              \and
              INAF/Istituto di Astrofisica Spaziale e Fisica Cosmica di Bologna,
              Via Gobetti 101, I-40129 Bologna, Italy
              \and
              Dip. di Fisica and INFN-Trieste,
              Via Valerio 2, I-34127 Trieste, Italy
              \and
              INFN-Pavia,
              Via Bassi 6, I-27100 Pavia, Italy
              \and
               Osservatorio Astronomico, Univ. di Perugia,
               Via B. Bonfigli, I-06126 Perugia, Italy
              \and
              ENEA-Bologna,
              via Martiri di Monte Sole 4 , I-40129 Bologna, Italy
              \and
              INFN-Roma \emph{La Sapienza},
              Piazzale A. Moro 2, I-00185 Roma, Italy
              \and
              INFN-Roma \emph{Tor Vergata},
              Via della Ricerca Scientifica 1, I-00133 Roma, Italy
              \and
              Dip. di Fisica, Univ. dell’ Insubria,
              Via Valleggio 11, I-22100 Como, Italy
              \and
              ENEA-Frascati,
              Via E. Fermi 45, I-00044 Frascati (Roma), Italy
              \and
               ASI, Viale Liegi 26 , I-00198 Roma, Italy
         }

   \date{Received: 16 August 2008; accepted 26 October 2008}

  \abstract
   {
     We report the results of a 3-weeks multi-wavelength campaign
     on the flat spectrum radio quasar 3C 273 carried out with the AGILE gamma-ray mission,
     covering the 30 MeV -50 GeV and 18-60 keV, the REM observatory (covering the near-IR and
     optical), Swift (near-UV/Optical, 0.2-10 keV and 15-50 keV), INTEGRAL (3 - 200 keV)
     and Rossi XTE (2-12 keV).
     This is the first observational campaign including gamma-ray
     data, after the last EGRET observations, more than 8 years ago.
   }
   {
     This campaign has been organized by the AGILE Team with the aim of observing,
     studying and modelling the broad band energy spectrum of the
     source, and its variability on a week timescale, testing the emission models
     describing the spectral energy distribution of this source.
   }
   {
     Our study was carried out using simultaneous light curves of the source flux
     from all the involved instruments, in the different energy ranges, in search for
     correlated variability. Then a time-resolved spectral energy distribution was
     used for a detailed physical modelling of the emission mechanisms.
        }
   {
     The source was detected in gamma-rays only in the second week
     of our campaign, with a flux comparable with the level
     detected by EGRET in June 1991. We found indication of
     a possible anti-correlation between the emission at
     gamma-rays and at soft and hard X-rays, supported by the
     complete set of instruments. Instead, optical data do not
     show short term variability, as expected for this source.
     Only in two precedent EGRET observations (in 1993 and 1997) 3C 273 showed intra-observation
     variability in gamma-rays. In the 1997 observation, flux variation in
     gamma-rays was associated with a synchrotron flare.

     The energy-density spectrum with almost simultaneous data,
     partially covers the regions of the synchrotron emission, the
     big blue bump, and the inverse-Compton.
     We adopted a leptonic model to explain the hard X/gamma-ray
     emissions, although from our analysis hadronic models cannot be ruled out.

     In the adopted model,
     the soft X-ray emission is consistent with combined
     synchrotron-self Compton and external Compton mechanisms,
     while hard X and gamma-ray emissions are compatible with  external Compton from
     thermal photons of the disk.
     Under this model, the time evolution of the spectral energy distribution
     is well interpreted and modelled in terms of an acceleration
     episode of the electrons population, leading to a shift in
     the inverse Compton peak towards higher energies.

   }
   {}

   \keywords{
     Gamma rays: observations --
     Galaxies: active --
     Galaxies: jets --
     Galaxies: quasar: general --
     Galaxies: quasar: individual: 3C 273 --
     Radiation mechanism: non-thermal
   }

\titlerunning{High Energy Variability of 3C 273 during the AGILE Dec. 2007 -
  Jan. 2008 campaign}

   \maketitle
%

\section{Introduction}

3C 273 is a very bright flat spectrum radio quasar. It is the nearest one (at a redshift of $z=0.158$),
and is a very peculiar AGN:
its spectral energy distribution shows the typical humps of blazars (\cite{padovani1995}), but
other features appear as well,
as the broad emission lines and the big blue bump typical of Seyfert-galaxies.

After 8 years from the last observations in gamma-rays, the AGILE mission
(\cite{agile}), with its GRID instrument (a pair conversion telescope, see \cite{grid} for details)
has opened new access to the observational window $30 \ MeV \div \ 50\ GeV$.\\
This source was discovered to emit in gamma-rays by COS-B in July 1976
(\cite{cosb_3c273}),
and observed again in June 1978 (\cite{bigne1981}); the mean flux detected
with COS-B was $\sim 60 \cdot 10^{-8} $ photons $cm^{-2}s^{-1}$ for E$>$100 MeV.
EGRET pointed this FSRQ several times, not always detecting it.
It was shown in gamma-ray activity in June
1991 (\cite{licti1995} reported a flux of $(56 \pm 8) \cdot 10^{-8}$ photons $cm^{-2}s^{-1}$ for E $>$ 70
MeV). In October-November 1993 \cite{monitigni1997} reported a flux variation
during the campaign from $(22\pm 5)\cdot 10^{-8}$ to $(56\pm 12)\cdot 10^{-8}$ photons $cm^{-2}s^{-1}$ for E $>$ 100 MeV).
The source showed large gamma-ray variability during a 7-weeks long EGRET
campaign (December 1996 - January 1997) (\cite{collmar2000} reported a
variation from $(25 \pm 9)\cdot 10^{-8}$ to $(76\pm 13)\cdot 10^{-8}$ photons $cm^{-2}s^{-1}$ for E $>$ 100 MeV), but no outstanding variation
was detected with COMPTEL (0.75-30 MeV).
During that campaign, a double synchrotron flare episode was detected by the UKIRT telescope of Hawaii observing in the near-IR (K band),
and by  RXTE/PCA (\cite{lawson1998}) in the 3-10 keV, showing correlated
variability  and $<1$ day lag of X-ray with respect to near-IR.
The flux variations (30-40\%) and the durations were similar at the two wavelengths.
In a joint X-ray and near-IR campaign in 1999, another flare was observed with a lag of 1 day
of X-rays with respect to near-IR. The flare lasted 2 days in the K-band
and 4 days in X-rays (\cite{sokolov2004}).

A multiwavelength campaign has been performed in June 2004 (\cite{turler2006}), triggered by
the sub-millimeter monitoring, observing a flux almost half than the lowest
jet activity ever observed (in a similar campaign of March 1986).
This campaign showed spectral features of the source usually overwhelmed by
the dominant jet activity. In particular the authors reported three further
weak humps located in the infrared, probably due to dust emission components.
%
%
%
%
%
%
%
%
%
%

Even if the spectral energy distribution of this source roughly shows the typical humps of blazars,
there is no general agreement on their origin. Not only the nature of the
hard X to gamma-ray emission is controversial, but also the big blue bump and the millimeter
to near-IR origin is in doubt.
In the June 1991 campaign, the low energy part of the spectrum showed a peak
at $6.7 \cdot 10^{11}$ Hz, and another not measured peak must be present in
$10^{13}-10^{14}$ Hz. With all these features, the theoretical modelling of
the SED is challenging.\\
Sometimes models for energy density distribution of blazars are loosely constrained, or
different models can be used to fit the same data. Studying the emission
evolution of the source (see \cite{boutelier2008} and references therein),
especially before, during and after flaring episodes in gamma-rays can help in
constraining the models.

In the case of 3C 273, the spectral energy distribution was studied
in very different theoretical scenarios (see for example \cite{monitigni1997}). The MeV peak has been fitted in the
context of pure synchrotron self-Compton, or in the context of the external Compton, considering
the photons of the big-blue bump (assumed to be emitted by the accretion disk)
as the seed photons for the inverse Compton.
Also proton-induced cascade models have been used, fitting the broad band energy spectrum
collected in Nov. - Dec. 1993 over more than 17 decades of energy.

We organized a 3-weeks multifrequency campaign on this bright source, with the
aim of building a simultaneous energy density distribution for each of the
3-weeks from near-IR to gamma-rays. In the following sections we report the
details of the observations, the data analysis and discuss the implication of
our results on the emission mechanisms of this source.
In the discussion of the spectral energy distribution,
we adopted a leptonic model to explain the hard X/gamma-ray
emissions, although our analysis cannot exclude  hadronic models.

%

\section{The multi-frequency observations}

%
We coordinated a multi-wavelength observational campaign to 3C 273
over 3-weeks, between 16 December 2007 and 8 January 2008.
The AGILE satellite pointed at the Virgo region for the entire period with its
gamma and hard X-ray instruments.
%
%
INTEGRAL  pointed at the source with
the complete set of its X- and soft gamma-ray instrumentation for
one complete revolution ($\sim$2.5 days) each of the 3-weeks.
Optical and near-infrared data were provided by the REM
observatory, that monitored the source every 2-3 days.\\
The source was found in high state at hard X-rays, and switched
from very low to intermediate/high-state in gamma-rays. Based on
our observations, we requested a Target of Opportunity (ToO)
observation for two pointings with the Swift observatory in the
last week of the campaign. The first Swift observation started 1.5
days after the end of the last INTEGRAL pointing.
Table \ref{tab:obs_schedule} summarizes the observations of the
campaign. A detailed description of the observations is given in
the next subsections.
\begin{table*}
  \caption{Schedule of the observations}
\label{tab:obs_schedule}
\centering
\begin{tabular}{l | c | c | c | l }
\hline\hline
Observatory & band/filter & start time (UT)    &  stop time (UT)       &  observing strategy \\ \hline
GRID        & 30 MeV - 50 GeV & 2007-12-16 17:14 & 2007-12-23 02:18 & nominal\\
            &                 & 2007-12-24 07:12 & 2007-12-30 23:03 & \\
            &                 & 2008-01-04 13:35 & 2008-01-08 11:06 & \\ \hline
SuperAGILE  & 18 - 60 keV     & 2007-12-16 17:14 & 2007-12-19 21:46 & nominal \\
            &                 & 2007-12-19 21:46 & 2007-12-23 02:18 & \\
            &                 & 2007-12-24 07:12 & 2007-12-27 15:07 & \\
            &                 & 2007-12-27 15:07 & 2007-12-30 23:03 & \\
            &                 & 2008-01-04 13:35 & 2008-01-08 11:06 & \\ \hline
JEM-X       & 3 - 35 keV      & 2007-12-19 18:08 & 2007-12-22 06:44 & rectangular dithering \\
            &                 & 2007-12-25 17:39 & 2007-12-28 06:27 & \\
            &                 & 2007-12-31 17:13 & 2008-01-03 04:00 & \\ \hline
ISGRI       & 18 - 400 keV    & 2007-12-19 18:08 & 2007-12-22 06:44 &  rectangular dithering \\
            &                 & 2007-12-25 17:39 & 2007-12-28 06:27 & \\
            &                 & 2007-12-31 17:13 & 2008-01-03 04:00 & \\ \hline
SPI         & 20 - 8000 keV   & 2007-12-19 18:08 & 2007-12-22 06:44 &  rectangular dithering \\
            &                 & 2007-12-25 17:39 & 2007-12-28 06:27 & \\
            &                 & 2007-12-31 17:13 & 2008-01-03 04:00 & \\ \hline
REM         & K           & 2007-12-11  8.20 & 2008-01-14  7.26   & every 2-3 days \\
            & H           &                     &                       &       \\
            & J           &                     &                       &       \\
            & I           &                     &                       &       \\
            & R           &                     &                       &       \\
            & V           &                     &                       & \\ \hline
UVOT        & V           & 2008-01-04 16:11 & 2008-01-04 17:47   & single exposure \\
            & B           &                     &                       &       \\
            & U           &                     &                       &       \\
            & UVW1        &                     &                       &       \\
            & UVM2        &                     &                       &       \\
            & UVW2        &                     &                       &       \\ \hline
UVOT        & V           & 2008-01-06 11:57 & 2008-01-06 15:24   & 3 exposures for each filter \\
            & B           &                     &                       &       \\
            & U           &                     &                       &       \\
            & UVW1        &                     &                       &       \\
            & UVM2        &                     &                       &       \\
            & UVW2        &                     &                       &       \\ \hline
XRT         & 0.2-10 keV  & 2008-01-04 16:11 & 2008-01-04 17:47  & PC + WT \\
            &             & 2008-01-06 11:57 & 2008-01-06 15:24  &  \\ \hline
%
%
\end{tabular}
\end{table*}

\subsection{AGILE observations}

The instrumentation carried by the Italian AGILE mission
(\cite{agile}) and used during the reported observations is composed by the
Gamma Ray Imaging Detector (GRID, 30 MeV - 50 GeV, \cite{presta}) and by the SuperAGILE
instrument (SA, 18-60 keV, \cite{superagile}). Both the
instruments perform simultaneous and co-aligned images over a
field of view in excess of one steradians, with a point spread
function (PSF) of $\sim$5$^{\circ}$/E(100 MeV) and 6 arcminutes,
respectively. Further details about the AGILE mission and the
individual instruments may be found in the cited papers.

AGILE monitored the source continuously from 2007-12-16 17:14 to
2008-01-08 11:06 UT, with two gaps of 1 and 4.5 days,
respectively, due to technical maintenance of the satellite. Both
GRID and SA were fully operational for the complete duration of
the observation. The resulting net exposure to 3C 273 for the GRID
and SA instruments was 742 ks for both.

Due to its solar panels constraints, the satellite bore-sight
drifts by $\sim 1^{\circ}$/day, and the target source drifts in
the field of view of the instruments consequently. During our
observation the source remained in the central $\pm 10^{\circ}$ of
both AGILE instruments for the whole campaign.

\subsection{INTEGRAL observations}

The INTEGRAL (\cite{wi03}) mission observed the source in the
revolutions 633 (from 2007-12-19 18:08 to 2007-12-22 06:44 UT),
635 (from 2007-12-25 17:39 to 2007-12-28 06:27 UT), 637 (from
2007-12-31 17:13 to 2008-01-03 04:00 UT) with the rectangular
dithering pointing strategy, for a total observing time of 7.5
days, corresponding to a net exposure to the source of
122 ks for JEM-X, 580 ks for ISGRI, and 494 ks for SPI.
The INTEGRAL observations are divided into
uninterrupted 2000-s intervals, the so-called science windows (SCWs).\\

The X-ray and soft gamma-ray observations were carried out with
JEM-X unit 1 in the range 3 - 35 keV (\cite{lun03}), ISGRI
(\cite{ube03}) in the range 18 - 400 keV, and SPI (\cite{varenne})
in the 20 - 8000 keV band.

\subsection{Swift observations}

The two pointings of the ToO Swift observation were carried out
between 2008-01-04 16:11 and 2008-01-04 17:47, and between
2008-01-06 11:57 and 2008-01-06 15:24. The first observation of
the Swift/X-Ray Telescope (XRT, see \cite{burrows2005} for
details), covering the 0.2-10 keV range, have an exposure of 454 s
in Windowed Timing (WT) mode and 2.5 ks in Photon Counting (PC)
mode; the second lasted for a total net exposure of 448 s
in WT mode and 2.8 ks in PC mode.
UVOT observed the source with all
lenticular filters except for the White one (V, B, U, UVW1, UVM2
and UVW2), with exposures of 213s for each optical
filters and 810, 610, 850s for the UV ones in the first observation;
268s for the optical filters and 537, 729, 358s in the UV for the second observation.\\

\subsection{All Sky Monitor and Burst Alert Telescope data}
For a continuous monitoring in the $2\ -\ 10$ keV energy band, we
retrieved the publicly available light curve data from the All Sky
Monitor\footnote{http://xte.mit.edu/asmlc/ASM.html} (ASM, \cite{levine}) onboard RossiXTE. \\
For the long term monitoring in the $15 \ - \ 50$ keV range we
downloaded the public light curve data for this source from the
Burst Alert Telescope\footnote{http://swift.gsfc.nasa.gov/docs/swift/results/transients/}
(BAT, \cite{bat}) onboard Swift.\\
Due to their observing strategy, both instruments provide sparse
observations of different durations. The typical exposure times
are 90 s for ASM and 840 s for BAT, and the
typical observation rate is 20 and 9
times per day, respectively.

\subsection{REM observations}

The near-infrared and optical monitoring was performed with the
Rapid Eye Mount (REM) telescope (\cite{zerbi2001}), for a  period
of 34 days, from 2007-12-11 08:20 UT to 2008-01-14  07:26 UT.

REM is a fully robotic, 60 cm telescope. It allows to execute
simultaneously optical and near-infrared photometry and
low-resolution spectroscopy. It hosts two parallel cameras: ROSS
(REM optical Slit-less Spectrograph) for optical observations
covering the range 0.45- 0.95 $\mu$m (V, R, I filters), REM-IR for
near-IR observations covering 0.95-2.3 $\mu$m range with 4 filters
(z, J, H and K). For this campaign we used the two instruments
with all their filters, except for the z on REM-IR, to obtain
nearly simultaneous data in order to study the almost
instantaneous spectrum of 3C 273.
The K, H, J images where exposed for 30 s, and the others for 300
s.
The sets of 6 bands observations were obtained every 2-3 days
during this 3-week campaign.


\section{Data Analysis and Results}

\begin{figure*}
\centerline{\psfig{figure=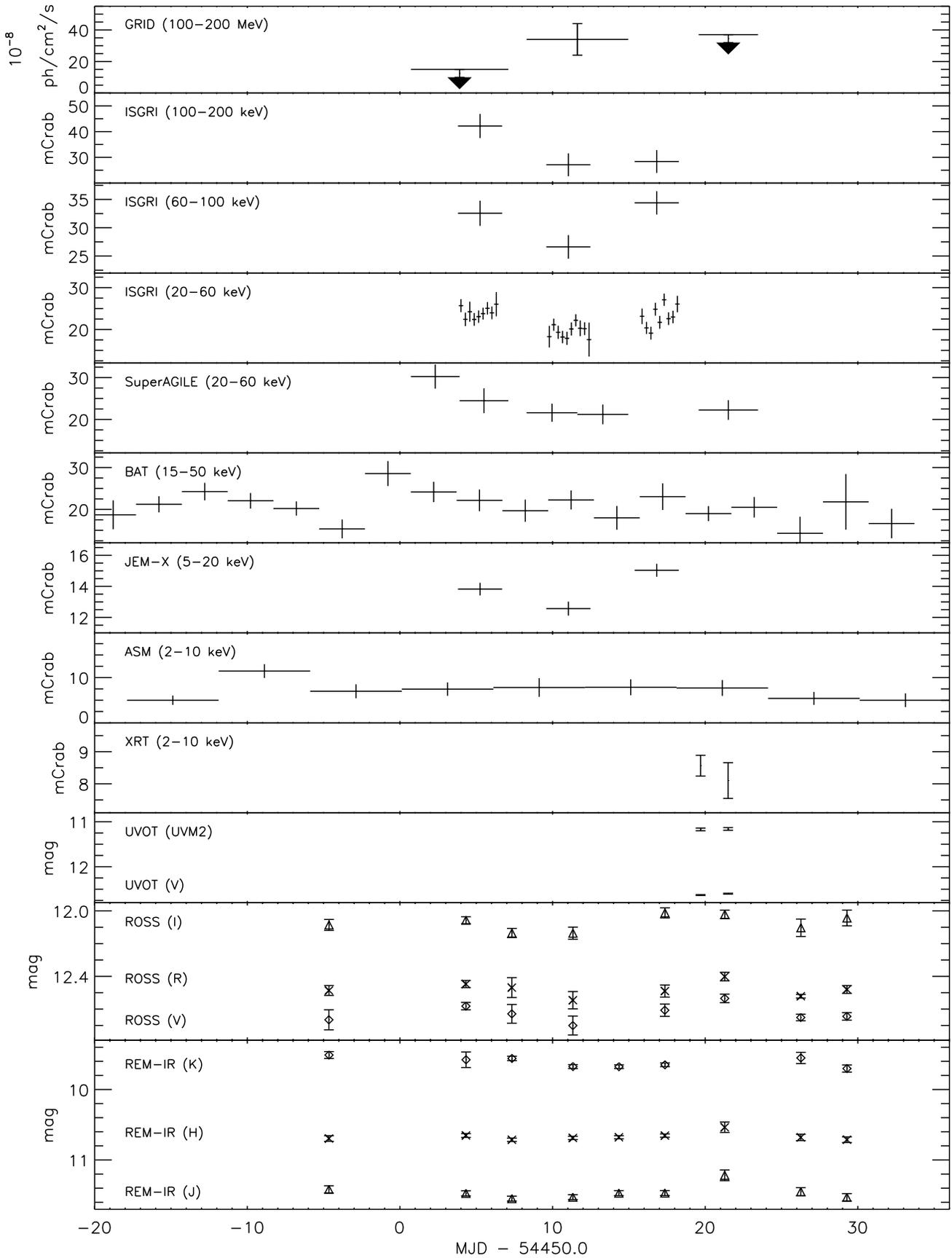,width=180.mm,height=230.mm}}
\caption{Complete set of multifrequency data collected during the AGILE observations of 3C 273.
From top to bottom: GRID data in the energy range
100-200 MeV, the ISGRI data in 100-200 keV, 60-100 keV, 20-60 keV,
SuperAGILE in 20-60 keV range, BAT in 15-50 keV, JEM-X in 5-20
keV, ASM in 2-10 keV, XRT in 2-10 keV, UVOT fluxes with UVM2 and V filters,
and REM fluxes with V (diamonds), R (crosses), I (triangles) filters from ROSS,
and J (triangles), H (crosses), K (diamonds) filters from REM-IR. The time is
referred to MJD 54450.0, corresponding to 2007-12-16 00:00:00 UT,
the starting day of our campaign.} \label{fig:lc_all}
\end{figure*}

%
The complete set of light curves from the multi-wavelength
campaign is shown in figure \ref{fig:lc_all}, ordered by
wavelengths (except for REM-IR data). The details of the data analysis from each instrument
are given below.

\subsection{AGILE GRID data}

Gamma-ray data from the GRID instrument were analyzed using the
standard AGILE/GRID pipeline (\emph{BUILD-15}).
The events taken
during the satellite passage in the south Atlantic anomaly were
rejected.
The GRID pipeline uses a Kalman filtering technique of the events
to identify the tracks, and to reconstruct the direction and the
energy of the incident gamma-rays.
To further reduce the charged particle tracks background, and to
select for the \emph{good quality} gamma-ray events, the Level-1
data were filtered  using \emph{FT3\_2}. This filter, based on multivariate
analysis, is the most selective one, concerning tracks selection
and quality factor for the accepted gamma-ray events.
Then Earth-albedo background was rejected,
excluding the gamma-rays produced inside a region of 10 degrees
from the Earth limb. GRID counts, exposure and Galactic background
maps were generated with a bin size of $0.3^{\circ} \ \times \
0.3^{\circ}$ for $E > 100$ MeV. The last step is
the AGILE Likelihood based analysis
(A. Chen et al. 2008, in preparation), running on these maps.
It is based on the point-source test statistic,
defined through the $T_S$ parameter
(this test statistic is described in detail in \cite{mattox} for the similar case of EGRET).
The $T_S$ parameter has the property that the $\sqrt{T_S}$ represents the
normalized significance of a detection.\\

Due to the continuous slewing of the satellite bore-sight, the
source moved in the field of view, but remained in the central
region (within $10^{\circ}$ from the on-axis) during the campaign.
This region has been well calibrated with the Vela pulsar
pointings during the science verification phase,
therefore the GRID flux estimate is corrected using the on-axis
calibration factor. We divided the total GRID observing time
in 3 blocks approximately one week long each.\\

The field of view in the proximity of the source was almost empty
for the first half of the observation. In the second half of the
observation, an unidentified source appeared at $\sim 5^{\circ}$
from 3C 273, rather bright in the last days of the observation.
Due to the presence of this unidentified source within a distance
comparable to the GRID PSF, the statistical uncertainties in the
estimation of the fluxes with the likelihood procedure are higher
than that of the previous period, causing the reduction of the
signal to noise ratio for 3C 273, mainly for the third observing block. \\

\begin{figure}
\resizebox{\hsize}{!}{\psfig{figure=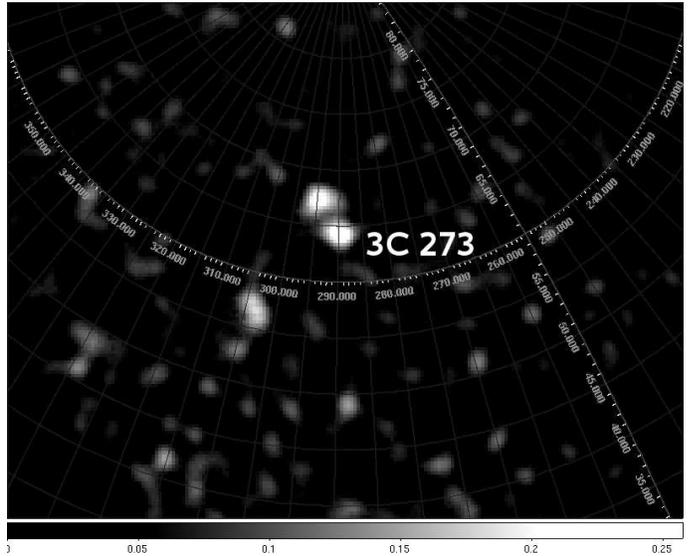}}
  \caption{
    Sky Image of Virgo region in gamma-rays obtained from GRID instrument onboard
    AGILE, from 2007-12-24 07:11:47 to 2007-12-30 23:03:06 UT, for E $>$ 100 MeV. The
    angular dimensions of the image is 60$^{\circ}$ width by
    45$^{\circ}$ height. The spot up/left near 3C 273 is the
    unidentified source, while the dimmer spot down/left is 3C
    279.
  }
  \label{fig:grid_sky}
\end{figure}

The sky image in the energy range 100 MeV - 50 GeV, exposed for the 7 central
days of the observation, is  shown in figure \ref{fig:grid_sky}.
The unidentified source is clearly visible in the image. 3C
279, the other well known blazar in the Virgo Region, appears very
faint, with a significance of 2.9 as measured by the
$T_S$ parameter (\cite{mattox}).\\

We found that by selecting the energy range 100 - 200 MeV we could
obtain a good rejection of the photons from the unidentified
source, still keeping the signal-to-noise ratio for 3C 273
unaffected. This suggests that the unidentified source has a very hard energy spectrum.\\

In the first and third week of our campaign 3C 273 was not
detected by the GRID, while in the second week it was detected at
a rather high gamma-ray activity, with a flux comparable to the
EGRET detection of the June 1991. The results of the analysis of
the GRID data is reported in table \ref{tab:grid_lc} for the three
individual blocks and for the whole period, both for the 100-200
MeV and $>$100 MeV energy bands. Upper limits with 95\% confidence
level are provided for the first and third week, when our analysis
provided flux estimations with $\sqrt{T_S} \ <$ 3 (\cite{mattox}).
The same data are also shown in the top panel of fig.
\ref{fig:lc_all}. As mentioned above, in the third observing block
the exposure lasted 4 days only, and the unidentified source was
very bright, thus the corresponding upper limits are higher with
respect to the first observing block.\\

\begin{table*}
\caption{3C 273 Flux measurements from the GRID instrument. For
the first and third block/week the $\sqrt{T_S}$ column provides the
value obtained by the standard processing.} \label{tab:grid_lc}
\centering
\begin{tabular}{l | cc | cc | cc | cc }
\hline\hline
energy & Flux during        & $\sqrt{T_S}$ &
Flux during       & $\sqrt{T_S}$ & Flux during
& $\sqrt{T_S}$ & Flux during            &
$\sqrt{T_S}$ \\
range  & observing block 1  &             & observing block 2 &             & observing block 3 &             & observing block 1+2+3  &             \\
(MeV)  & $(10^{-8} \ \gamma/cm^2/s)$ &  & $(10^{-8} \ \gamma/cm^2/s)$  &  &
$(10^{-8} \ \gamma/cm^2/s)$ &  & $(10^{-8} \ \gamma/cm^2/s)$ &  \\
\hline
$100 \div 200$ & $<15$ & 0.2  & $34_{-10}^{+12}$ & 4.6 & $<37$ & 0.9 & $17 \pm 6$ & 3.8 \\
$>100$         & $<20$ & 1.4  & $33 \pm 11$     & 4.4 & $<50$ & 1.5 & $22 \pm 6$ & 4.6 \\
\hline
\end{tabular}
\end{table*}

\subsection{SuperAGILE data}

The AGILE/SuperAGILE instrument (\cite{superagile}) provided
images of 3C 273 in the energy range 18-60 keV, during the same
period of the GRID. Based on the available statistics, we divided
the complete 3-week observation in 5 bins of 3-4 days each. The
first two SA bins are simultaneous to the first GRID block, the
next two bins are simultaneous to the second GRID block, while the
last SA bin is simultaneous to the third GRID block, due to the
shorter exposure in the third week.\\

\begin{figure}
  \resizebox{\hsize}{!}{\psfig{figure=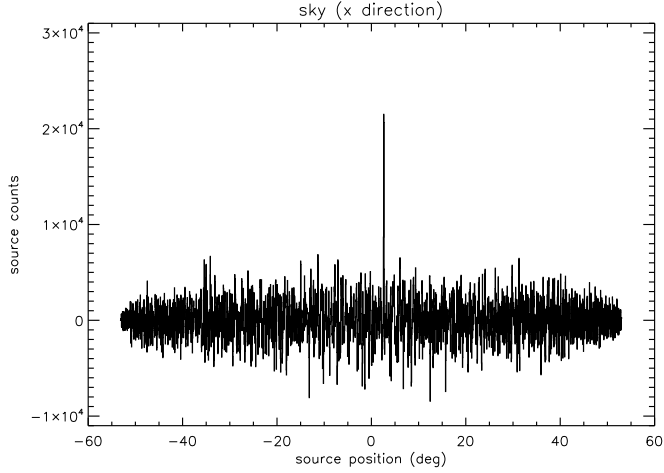}}
\caption{
  One-dimensional sky image of Virgo field in hard X obtained from SuperAGILE instrument
  onboard   AGILE, where only 3C 273 shows up as a peak. Data were integrated from
  2007-12-24 07:11:47 to 2007-12-30 23:03:06 UT in the 20-60 keV energy range.
  Abscissa reports the angular distance (degrees) from the on-axis direction.
}
\label{fig:sax_sky}
\end{figure}

The SA instrument is a one-dimensional coded-mask imager,
producing two orthogonal one-dimensional sky images of the
observed sky, starting from photon-by-photon data on user-defined
time intervals. As an example, the sky image provided for one of
the two instrumental coordinate (the \emph{X} coding direction)
accumulated over the third and fourth time-bins
is shown in fig. \ref{fig:sax_sky}.\\

The SA data analysis was performed with the \emph{TDS source}
package of the \emph{SASOA} pipeline (\emph{build 3.8.0}). The
photon lists are filtered, excluding those events taken during the
satellite passage through the south Atlantic anomaly, and for
events taken when the source was occulted by the Earth.\\

The continuous slew of the pointing direction of 1 degree/day due
to the AGILE solar panel constraints, requires that the detector
images accumulated from the event list are corrected using
pointing information from the star-tracker data.\\
Dealing with a coded-mask imager, the attitude-correction depends on the source
position in the field of view of each detector.
Anyway, the correction to apply changes slowly with the source position in the
FOV. Thence the correction calculated for a specific position
$\big[\theta^*_{cod},\theta^*_{uncod}\big]$ in the FOV
(where $\theta^*_{cod}$ and $\theta^*_{uncod}$ represent the positions
in the coded and uncoded direction respectively)
can be applied without affecting the point spread function of sources
located at some degree from $\big[\theta^*_{cod},\theta^*_{uncod}\big]$. On account of this,
we calculate the correction in a grid of 19x17 positions in the FOV, with
a grid step of 6$^{\circ}$ along the detector coded direction, and 4$^{\circ}$ along the non-coded
direction. Thence $19 \times 17$ virtual detector images are generated from the photon list of each
detector. A detailed description of the attitude correction procedure for
SuperAGILE will be presented in a forthcoming paper (Pacciani et al. 2009, in preparation).
A cross-correlation procedure of these detector images with the mask code
provides the images of the point like sources, as shown in fig. \ref{fig:sax_sky}.\\


The count rate of photons collected from each silicon $\mu$strip of the detectors
is affected by the non-uniformities between the energy thresholds of the
analog chains (see \cite{pacciani_thr}), and to the temperature dependance of the discriminator
units. we account for this non-uniformity applyng a detector efficiency vector
in the imaging procedures.
%
The efficiency is generated from a
blank field and corrected for the temperature effects. The
non-uniformities in the low-energy thresholds makes it unsafe to
use the 18-20 keV energy bin for long integrations, when threshold
variations can critically affect the results. For our analysis we
then used SA data in the 20-60 keV energy range.\\

The SA response was calibrated in-flight with a raster scan with
the Crab Nebula, at several positions in the FOV. During our
observation 3C273 scanned the central part of the FOV, ranging
from 7.7 to -2.4 deg in the X instrumental coordinate and
from 11.0 to -12.4 deg in the Z instrumental coordinate.
Observed count rates are then converted into physical units of
mCrab, by using the Crab response at the relevant position in the
FOV (implicitly assuming a Crab-like energy spectrum). The average
20-60 keV flux measured by SA over the complete 3-week observation
is (23.9 $\pm$ 1.2) mCrab, with a source detection
significance of 14 $\sigma$ and 16 $\sigma$ in the X and Z
coordinate, respectively and a net exposure to the source of
742 ks. The results of a time-resolved analysis are
reported in the relevant panel of fig. \ref{fig:lc_all} in the
20-60 keV energy range, adding up the normalized count rates from
each one-dimensional sky image. The corresponding data are
reported in the appendix, table \ref{tab:hardx_lc_app}.

\subsection{INTEGRAL JEM-X, ISGRI and SPI data}
Wide-band data for the source were obtained using the high-energy
instruments onboard INTEGRAL, JEM-X in the effective energy range
5-20 keV, ISGRI in 18-200 keV and SPI in 100-500 keV.
The effective energy ranges we used exclude the energy regions with too low effective
area and the lowest energies, affected by electronic noise.
For SPI we reported the energy range where the effective area is comparable or higher than
the ISGRI.
Data were
processed using the Off-line Scientific Analysis OSA 7.0 software
released by the Integral Scientific Data Centre. ISGRI light
curves and spectra were extracted for each individual SCW. The
spectrum from JEM-X was extracted from a mosaic image at the
position of the source. Due to the dithering pointing strategy,
the source is not always in the JEM-X field of view. The SPI data
were integrated for the three INTEGRAL revolutions together to
achieve the needed sensitivity up to 500 keV. The net exposure to
the source was 122 ks for JEM-X, 580 ks for ISGRI
and 494 ks for SPI. The average measured flux
for each instrument was: (13.81$\pm$0.25) mCrab
in 5-20 keV (JEM-X, 56 $\sigma$ detection significance),
(22.30$\pm$0.32) mCrab in 20-60 keV (ISGRI,
70 $\sigma$ detection significance), and
(41$\pm$9) mCrab in 100-500 keV (SPI, 4.4 $\sigma$
detection significance) which provides a marginal detection.\\
The light curves in the energy ranges 5-20, 20-60, 60-100, 100-200
keV from the above instruments are shown in figure
\ref{fig:lc_all} with a bin size of 200 ks (an INTEGRAL
revolution), except for the 20-60 keV energy range, where the
counting statistics allowed for a 25 ks bin size. The data are
reported in table \ref{tab:hardx_lc_app} of the appendix.
The simultaneous 20-60 keV flux measurements by SuperAGILE and ISGRI
appear in good agreement.\\

The spectra taken during the three individual INTEGRAL
revolutions, can be fitted with a simple power law model in the
18-120 keV energy range. The best-fit parameters are reported in
table \ref{tab:he_isgri_fitting}. No significant spectral
evolution is detected, except for a marginal evidence of softening
in the spectrum from revolution 635.

\begin{table}
\caption{Spectral fitting parameters in the 18-120 keV energy
range (uncertainties at 90\% level).} \label{tab:he_isgri_fitting} \centering
\begin{tabular}{cccc}
\cline{2-4}  \cline{2-4}
                 & $1^{st}$ week     & $2^{Nd}$ week      & $3^{rd}$ week \\
                 & (rev 633)         & (rev 635)         & (rev 637)    \\ \hline
photon index     & $ 1.77 \pm 0.07$  & $ 1.87 \pm 0.09 $ & $ 1.80 \pm 0.07 $ \\
flux (20-40 keV) &  $173 \pm 5$      &  $144 \pm 5$      & $169 \pm 5$   \\
10$^{-12}$ erg cm$^{-2}\ s^{-1}$  &    &                   &  \\
\hline
\end{tabular}
\end{table}

\subsection{Hard X-ray data from BAT}

Flux measurements of the sources serendipitously observed by the
BAT instrument onboard Swift are available on-line for every
satellite orbit. The flux measurements are sparse and with
different exposure, depending on the specific satellite pointing
strategy. We grouped the available data with bin size of 3 days.
To account for the huge spread in the signal to noise ratio
between data, a weighting factor inversely proportional to the
flux error was applied during the rebinning operation.\\

The BAT light curve in the range $15 \ - \ 50 $ keV is shown in
figure \ref{fig:lc_all} and reported in table
\ref{tab:hardx_lc_app} of the appendix. The light curve from BAT
has the same trend as the SuperAGILE and ISGRI instruments but a
slightly lower flux, likely due to the slightly different
bandpass.\\

\subsection{XRT data}
For the analysis of soft X-ray data from the Swift X-Ray Telescope
(XRT), we used the version 11.6 of the XRT
pipeline\footnote{http://heasarc.nasa.gov/docs/swift/analysis/xrt\_swguide\_v1\_2.pdf.}.
Grade filtering was applied by selecting the 0-2 and 0-12 ranges for the
data collected in WT and PC mode, respectively. The data collected in PC mode are affected
by pile-up in both observational epochs (average count rate
$\sim$8 counts/s). The pile-up estimation and correction was made
more difficult by the presence of a bad column crossing the center
of the source extraction region in both observational epochs.
Thus, we could obtain only a rough estimation of the pile-up
effects and decided to use only the data collected in Windowed Timing
mode, not affected by pile-up.\\

To account for the bad column in the light curve and spectra
extraction, we used the exposure maps computed for each epoch and
from them we generated the ancillary response files. The latter
are very sensitive to the source centroid position on the CCD. But
due to the bad column, we could not evaluate the centroid
accurately and therefore run the pipeline fixing the source
position at the coordinates given from optical and radio
observations in the SIMBAD archive. The Swift star sensors
precision introduces a systematic uncertainty in the evaluation of
the satellite pointing, providing a mismatch between the source
centroid on the CCD evaluated with the star sensors data and the
effective one. In order to evaluate the effects of this
systematics on the flux and spectral index estimation, we computed
the effective area also over two other positions shifted of 3.1"
(a region that encloses 90\% of the Point Spread Function from the SIMBAD one).\\
The signal was extracted from a rectangular region (40 pixels wide
and 20 pixels in height), assuming as nominal the position
centered on the SIMBAD coordinates. The difference between the
results obtained at the SIMBAD position and the shifted ones is
then taken as a systematic uncertainty, denoted below as "(syst)".
Assuming an absorbed simple power law spectral model, with
absorption fixed at $N_H=1.79 \times10^{20}$ cm$^{-2}$ (Kalberla
et al. 2005), we found a photon index of $1.61\pm0.05$, with an
observed 2-10 keV flux of $1.85\times10^{-10} \ \pm \ 0.04 \
(stat) \ \pm 0.03 \ (syst) $ erg cm$^{-2}$ s$^{-1}$ during the
first epoch (reduced $\chi^2$ is 0.9, 92 d.o.f.). No
significant variations were observed during the second epoch,
where the photon index is $1.57\pm 0.06$ and the observed 2-10 keV
flux is $1.75\times10^{-10}   \ \pm \ 0.04 \ (stat) \ \pm 0.08 \
(syst)$ erg cm$^{-2}$ s$^{-1}$ (reduced $\chi^2$ is 1.0, 71 d.o.f.).
The bad quality of the image in the second observation caused the
systematics to be higher. The star sensor systematics does not
affect the photon index estimation in WT mode.\\
XRT data are shown in figure \ref{fig:lc_all}.

\subsection{UVOT data}
UV data reduction and photometry of the source was performed using
the standard UVOT software developed and distributed within the
HEAsoft 6.3.2 by the NASA/HEASARC and the most recent
calibrations included in the last release (2007-07-11) of the
``Calibration Database'' (CALDB; see also \cite{Poole08}). Source
counts were extracted for all filters from circular aperture of
5\arcsec\, radius, the background from source-free circular
aperture of 12\arcsec\ radius and count-rates converted to fluxes
using the standard zero points. The count-rate of the source is
near the limit of acceptability for the ``coincidence loss''
correction factor included in the CALDB ($\sim$\,90 {\rm cts
s}$^{-1}$), in filters U, B, UVW1 and UVW2 for both observations.
We considered in our analysis only the V and UVM2 filters for both
the observations and also the B for the second. The fluxes were
then de-reddened using a value for $E(B-V)$ of 0.021 mag
(\cite{Schlegel1998}) with $A_{\lambda}/E(B-V)$ ratios calculated
for UVOT filters (for the latest effective wavelengths) using the
mean Galactic interstellar extinction curve from
\cite{Fitzpatrick1999}. No significant variability was detected within
each single exposure for both the observations.\\
UVOT data are shown in figure \ref{fig:lc_all} and reported in table \ref{tabapp:rem_lc_app} of the
appendix.

\subsection{REM data}
Data reduction and photometry of the near-IR and optical frames
from the REM observations has been carried out through the GAIA
\footnote{
http://docs.jach.hawaii.edu/star/sun214.htx/sun214.html} software
using images corrected by bias, dark and flat-field (see
\cite{stetson1986}). The instrumental magnitudes have been
calibrated using the comparison star sequences reported in
\cite{gonzales} for the optical and the near-IR bands. Three
bright isolated stars in the field of view were used as reference
to calculate the instrumental magnitude shift.\\
The near infrared and optical light curves over a 34 days
monitoring for the K, H, J, I, R, V bands of \emph{REM}
observatory are shown in figure \ref{fig:lc_all} and reported in
table \ref{tabapp:rem_lc_app} of the appendix. The large errors
for some data point are due to the presence of the moon, causing
errors in the photometry of 3C 273 and/or of the reference stars.
Small differences in the simultaneous measurements in the UVOT/V
and REM/V bands are most likely to be ascribed to the slightly
different bandpass, as reported in table \ref{tabapp:rem_lc_app}.

\section{Discussion}
%
From the multi-frequency light curves shown in fig.
\ref{fig:lc_all}, the source exhibited gamma-ray activity in the
second week of the AGILE observation. In the same time period, a
$\sim20\%$ reduction in the X- and hard X-ray flux was detected by
all the involved instruments. Instead, the near-IR, optical and UV
fluxes remained constant to within $\sim10\%$ variability.
No strong evidence for correlations can then be derived between the
gamma-ray activity and the source emission in other bands,
using the analysis of the light curves.

In order to study the spectral variability, in the following we
divided the campaign in 3-weeks, according to the GRID
observing blocks. We first evaluated the possible contribution of a
Seyfert-like reflection component in the X-ray/soft gamma-ray
energy spectrum, and then build the complete Spectral Energy
Distribution for two epochs, to understand the origin of the
gamma-ray activity.
%

\subsection{Limits on Seyfert-like spectral features}
The wide band energy spectra of the source taken by BeppoSAX
between 1997 and 2000 allowed to disentangle the contribution of
the jet and Seyfert-like features (see \cite{grandi2004}). The XRT
calibration status below 0.6 keV (see \cite{cusumano2007}) does
not allow us to study the soft excess, while the Iron line studies
(\cite{yaqoob2000} and references therein) are prevented to us by
the counting statistics. Instead, our data allow us to study the
reflection hump contribution to the spectra, emerging at 20-60 keV.\\

Unfortunately, results in this spectral region are very sensitive
to possible uncertainties in the cross-calibrations between JEM-X
and ISGRI instruments. In the $3^{rd}$ IACHEC meeting\footnote{
http://www.iachec.org/iachec\_2008\_meeting.html} (held in Schloss
Ringberg, Germany May 18-21 2008), cross-calibration factors near
to unity were reported for the instruments onboard INTEGRAL (see
the J. P. Roques presentation\footnote{
http://www.iachec.org/2008\_Presentations/Roques\_SPI.pdf}) for
the Crab observations. In the following we use that cross
calibration factors for the instruments onboard INTEGRAL, and keep
free the XRT normalization factor (also to account for the
systematics in XRT data relative to our specific observation). It
is also important to note here that the declared INTEGRAL
cross-calibration factors are reported for Crab-like spectra,
while the energy spectrum of 3C 273 is harder (table \ref{tab:he_isgri_fitting}). In order
to account for possible spectral dependence of the cross-calibration
constants, we always kept the JEM-X factor fixed (to 1.02) and
fixed the ISGRI constant to 3 possible values, reporting the best-fit
results in all the three cases. SPI data were not used in this analysis.\\

We built three energy spectra (one per INTEGRAL revolution). Each
spectrum contains the ISGRI data for that revolution. The first
XRT observation was performed 1.5 days after the end of the last
INTEGRAL pointing, thence we used XRT data for the third spectrum
only. In order to reach enough significance, all the JEM-X data of
the campaign were merged together in the spectrum of the third
week. We corrected the JEM-X multiplicative factor to account for
the true normalization factor for the third week (during the third
week the JEM-X flux was 1.08 times the mean flux of the
campaign).\\

The energy spectra of the three epochs were fitted simultaneously.
We first attempted a fit with an absorbed simple power law plus a
Compton reflection hump described by the PEXRAV model in the XSPEC
package (\cite{pexrav}). We used the PEXRAV parameters set
proposed in \cite{grandi2004}, with only the PEXRAV normalization
allowed to vary in the fitting, but linked for the three epochs.
The photoelectric absorption column was fixed to the Galactic
value of $N_H \ = \ 1.79 \cdot 10^{20}$ cm$^{-2}$ (\cite{dickey}).
The power law parameters were linked for the three epochs, except
for their normalization, left completely free to vary. With this
approach, we tested the hypothesis that the hard X-ray variability
among the three epochs was entirely due to the jet-component. The
best-fit result is marginally acceptable ($\chi^2/d.o.f.$=1.15, 47
d.o.f., null hypothesis probability 0.24).\\

We then introduced a break in the description of the jet component
(that is, we introduced a broken power law in place of the simple
power law) and adopted the same fitting strategy, again under the
hypothesis of a variability entirely due to the jet component. An
acceptable fit was achieved, and the best fit results are reported
in the first column of table \ref{tab:pexrav}, where uncertainties
on parameters are computed at 90\% for one interesting parameter.
We note that using only a broken power law without a reflection
component provides a significantly worse best-fit result, with a
$\chi^2/d.o.f.$ of 1.20 (46 d.o.f., null hypothesis probability 0.17).
Interestingly, the fit
would become fully acceptable if the JEM-X/ISGRI cross-calibration
is allowed to go in the range
$C_{ISGRI}/C_{JEM-X}=1.25 \div 1.30$.\\

To the aim of providing the reader with the confidence on how
strong the need for a Compton reflection component is in our
spectra, we also studied the case where the difference in spectrum
between Crab and 3C 273 may bring to a different cross-calibration
factor between JEM-X and ISGRI. We tested the cases of
$C_{ISGRI}=0.89$ and $C_{ISGRI}=1.09$, and the best-fit parameters
are given in columns 2 and 3 of table \ref{tab:pexrav}. As
expected from our previous discussion, the higher the ISGRI/JEM-X
cross-calibration factor is, the lower is the needed contribution
by the Compton reflection. But a minimum value of 1.25 is needed
to exclude it, and this contrasts with the
latest releases by the hardware teams.\\

\begin{figure}
\resizebox{\hsize}{!}{\psfig{figure=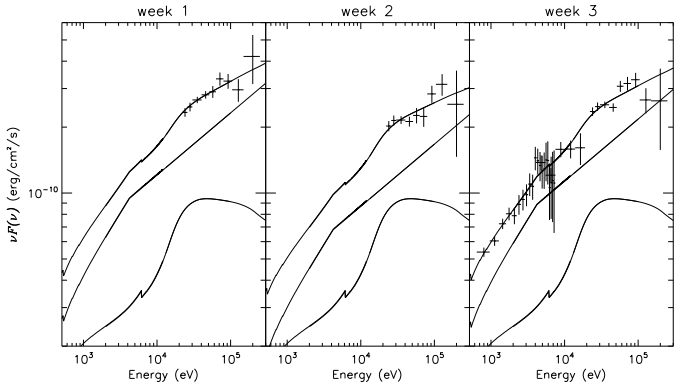,width=150.mm}}
\caption{
  Hard X spectral energy density obtained with ISGRI data for the 3-weeks of the campaign.
  Each column refers to a week.
  The SED of the third week includes the JEM-X and XRT data.
  The overplotted curves refer to the Compton reflection hump (the lower curve), the broken
  power law (the middle one), and the summation of the two (upper curve).
  We used the INTEGRAL Cross calibrations factors obtained for the Crab.
}
\label{fig:sed_hardx}
\end{figure}

We note that the uncertainties on the
normalization factor for the PEXRAV and broken power law 
(indicated respectively as $N_{Pex}$ and $N_{Bkp}$ in table \ref{tab:pexrav})
are correlated.
Therefore, in order to compare the
contribution of the jet in the three epochs, here in terms of the
value of the normalization of the broken power law, we performed
another fit by fixing the PEXRAV parameters to their best fit
value.
The uncertainty on the $N_{Bkp}$ under this assumption are
provided in parenthesis in column 1 of table \ref{tab:pexrav},
showing that the jet-component variation between the first and the
second week is indeed statistically significant, while the
difference between the values in the second and third is
marginally consistent with the combined 90\% uncertainties on the
individual parameters. 
The spectral energy densities for each week
with the best-fit models are shown in fig. \ref{fig:sed_hardx}.
The reflection hump discussed in the previous section is not included in the model.

Thus, from our analysis of the time-resolved X-to-soft-gamma-ray
energy spectrum, we can derive that the variability observed from
the light curves in this energy range is most likely due to the
jet component, described as a broken power law in our emission
model, although a non-variable reflection component is also
required by the spectral data presented here.

\begin{table*}
\caption{Best-fit results for the simultaneous spectral fit of the
three epochs (see text for details). We report the values obtained
for the \emph{nominal} Crab cross-calibration factor of ISGRI
($C_{ISGRI}=0.99$), and for $C_{ISGRI}=0.89$ and $C_{ISGRI}=1.09$.
$N_{Pex}$ and $N_{Bkp}$ are the normalization factors for the PEXRAV and
broken power law models respectively (reported as photon flux at 1 keV in units of $10^{-3} \ ph/cm^2/s/keV$).
 $\ \ \ ^*$ error obtained fixing the PEXRAV normalization.}
\label{tab:pexrav} \centering
\begin{tabular}{lc|c|c}
\cline{2-4}\cline{2-4}
                     & $C_{ISGRI}=0.99$    & $C_{ISGRI}=0.89$  & $C_{ISGRI}=1.09$ \\
                     & (nominal from Crab)&                  &  \\ \hline
$N_{Pex}$ ($10^{-3} \ ph/cm^2/s/keV$) & $13.3 \pm 5.0$ ($\pm 0$)$^*$ & $19.4 \pm 5.3$ & $7.9 \pm 4.7$\\
$N_{Bkp}$ for rev. 633 ($10^{-3} \ ph/cm^2/s/keV$)  & $27.1 \pm 9.3$ ($\pm 5.1$)$^*$& $20.3 \pm 9.5$ & $33.8 \pm 9.3$\\
$N_{Bkp}$ for rev. 635 ($10^{-3} \ ph/cm^2/s/keV$)  & $19.5 \pm 7.9$ ($\pm 3.8$)$^*$& $13.5 \pm 7.8$ & $25.7 \pm 8.0$\\
$N_{Bkp}$ for rev. 637 ($10^{-3} \ ph/cm^2/s/keV$)  & $25.4 \pm 8.6$ ($\pm 4.7$)$^*$& $18.7 \pm 8.7$ & $32.0 \pm 8.6$\\
photon index 1       & $1.46 \pm 0.12$ ($\pm 0.10$)$^*$& $1.38 \pm 0.15$& $1.51 \pm 0.10$ \\
photon index 2       & $1.71 \pm 0.05$ ($\pm 0.04$)$^*$& $1.66 \pm 0.05$& $1.75 \pm 0.04$\\
break Energy (keV)   & $4.   \pm 2.$ ($\pm 2.$)$^*$& $4.   \pm 2.$ & $4.   \pm 2.$\\
XRT cross-calib      & $1.01 \pm 0.11$ ($\pm 0.11$)$^*$& $1.02 \pm 0.11$ & $0.98 \pm 0.11$  \\
$\chi^2/d.o.f.$      & 38.0/45 & 40.7/45& 36.4/45\\
null hypothesis probability    & 0.76    & 0.65  & 0.82 \\
\hline
\end{tabular}
\end{table*}

\subsection{Spectral Energy Distribution}

With the aim of understanding the origin of the gamma-ray
emission, we used our multi-frequency data to build a Spectral
Energy Distribution (SED). Due to the uncertainty in the
evaluation of the gamma-ray flux of 3C 273 for the third week, in
the following we refer mainly to the first and the second week of
observations.\\

We made the following approximations in the evaluation of the SED.
Similarly to the case discussed in the previous section, for JEM-X
we used the spectrum extracted from all the three JEM-X
observations together, and we applied a correction factor to the
spectra to obtain the observed count rate (in the 5-20 keV band)
from each revolution. Due to the statistics, the SPI data are
obtained from the integration of the three INTEGRAL revolutions
together. Finally, we assumed a photon index of 2.4 to convert
counts to photons in the AGILE GRID data.

\begin{figure}
\centering
\begin{tabular}{c}
\resizebox{\hsize}{!}{\psfig{figure=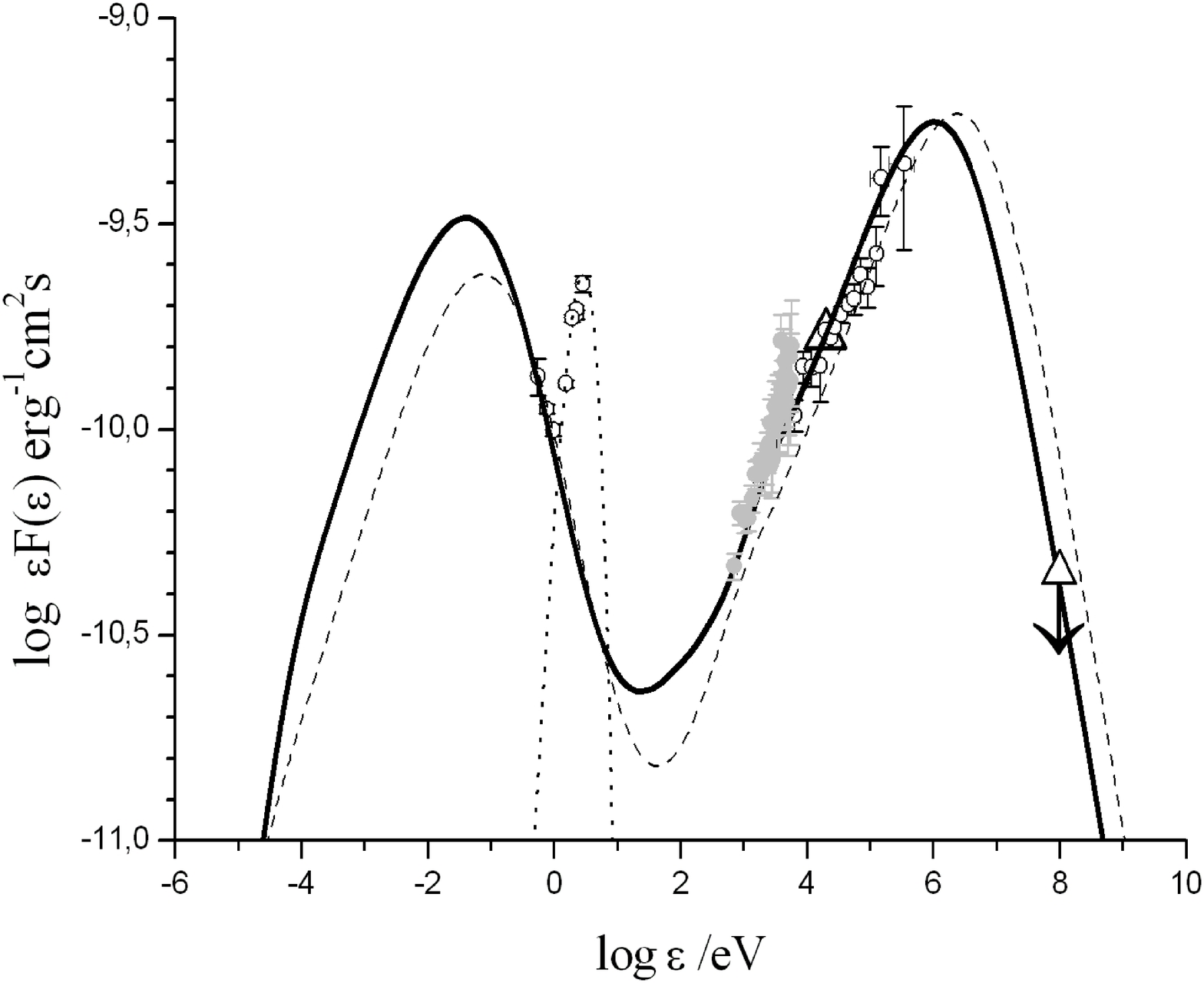}} \\
\resizebox{\hsize}{!}{\psfig{figure=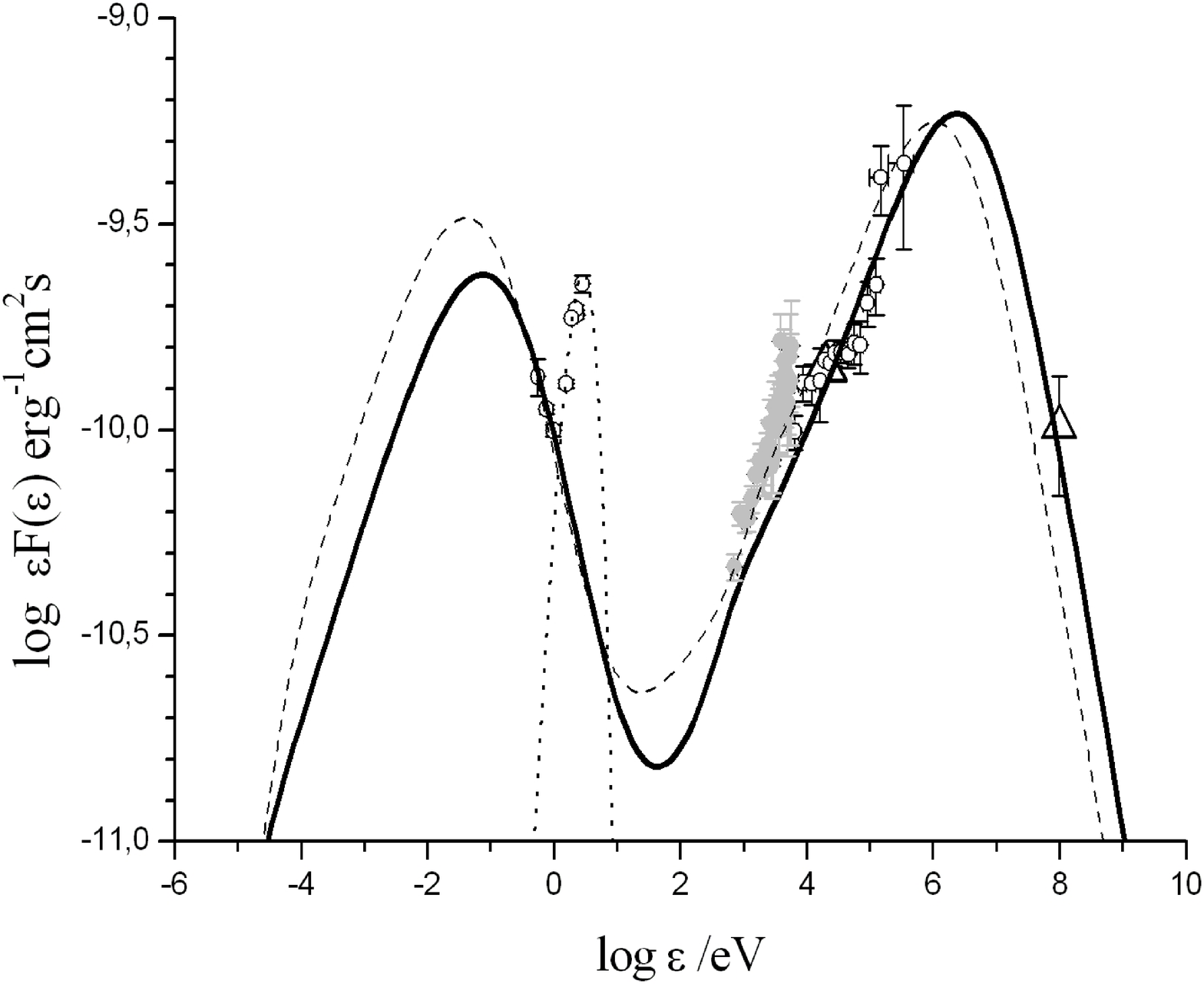}}
\end{tabular}
\caption{
  Spectral Energy Distribution of 3C 273 for the first (top panel) and
  the second week (bottom panel). Triangles are for AGILE data. The grey data refers to the XRT
  observations, performed in the third week.
  The line is the model for the simultaneous data of the week.
  The model for the other week is reported for comparison as dashed
  line. The reflection hump is not included in the model.
  Where not visible the energy range is smaller than the symbol.
}
\label{fig:sed}
\end{figure}

The resulting SEDs for the first and second week are shown in
fig. \ref{fig:sed}. We described the broad band emission in the
framework of a model including synchrotron emission, synchrotron
self Compton and external Compton components (see
\cite{maraschi1992}, \cite{marscher1992}, \cite{sikora1994}).
We didn't take into account the reflection hump in the SED model.\\
Remarkably, the flux distribution in our high gamma-ray state
period is similar to that measured during the multi-wavelength
campaign performed in June 1991, when gamma-ray variability was
not observed. In that campaign (\cite{licti1995}) the gamma-ray
flux was $(56 \ \pm \ 8)\cdot 10^{-8} $  photon $ cm^{-2} s^{-1} $ for E $>$ 70 MeV,
and the photon index $2.39 \ \pm 0.13$, consistent with the AGILE
flux of $\big(34_{-10}^{+12}\big)\cdot 10^{-8}$ $ \ photon \ cm^{-2} s^{-1} $ for E $>$ 100 MeV.

The observed variability of the SED between the two epochs cannot
be associated to a synchrotron flare. In that case an enhancement
of the emission at all the observed wavelengths is expected. The
variability behaviour can be reproduced as a shift toward higher
energies of the electron density, thence related not to 
the injection of a new blob, but to electron acceleration. According
to this hypothesis, we modelled the variability keeping the bulk
Doppler factor, the blob radius and the disk luminosity unchanged.
Instead, we varied the parameters related to the accelerated electrons: i.e. the
electrons energy distribution ($n_e$, $\gamma^*$ and $p_2$) and, slightly, the tangled
magnetic field.
But the choice of the SED parameters allowing for a change from the
first to the second week is not unique.
The chosen parameters of the SED model for the two epochs
are reported in table \ref{tab:sed}.\\

Actually, the spectral variability that we observed can be
interpreted in the context of standard model of FSRQ as follows.
The flux at frequencies $>3 \cdot 10^{14}$ Hz, consistently with
the large ($6^{\circ}$) viewing angle, appears dominated by
thermal emission from the disk and/or from the BLR. Thus, we
expect the emission in the range of frequencies observed by REM
to not vary on daily timescales, and to hide variations of the
synchrotron emission except in the near-IR (K and H bands).
The REM observations, show variations
lower than $\sim 10-15\%$ in the near-IR and optical.
%
%
Our model, showed in fig. \ref{fig:sed},
produce no variations of the synchrotron emission
in the near-IR and optical energy regions.\\
A moderate shift of the direct synchrotron spectrum towards higher
frequencies is detectable in the far-IR and in the soft X (if not
hidden by other thermal components, e.g. the components suggested in
\cite{turler2006}, and the soft excess reported in \cite{grandi2004}).
But we didn't have coverage of that energy regions for all the campaign.\\

Variations are instead revealed in the inverse
Compton reprocessing in the X-ray and gamma-ray domain. The
relative variations that we detected, $\sim$20-30$\%$ and a factor
$\sim$2-3, respectively, together with the fact that the gamma-ray
flux appears anti-correlated to the X-ray flux, indicates that a
shift toward higher energy in the electron density is very likely
responsible for the observed variability.

In the model, the associated SSC variation reflects in a moderate decrease of the leading edge of the SED in
the soft X-ray band, whereas the EC by the disk shows up as a flux
decrease in the hard X-rays. In the gamma-ray band, the falling
portion of the EC spectral energy distribution well describes the
observed enhancement.

%

%
\begin{table*}
\caption{Parameters for the spectral energy distribution for the first and
  second week of the campaign.
  $p_1$ and $p_2$ are the pre and post break spectral index for the electron
  population, $\gamma^*$ is the break energy Lorentz factor, $\gamma_{min}$ is
  the cut-off energy of the electron population, B the tangled magnetic field, $r$ the
  radius of the spherical blob in the comoving frame, $\delta$ the Doppler
  factor, $n_e$ the electron density. 
}
\label{tab:sed}
\centering
\begin{tabular}{ l l l c c c c r c c }
\hline \hline
week   & $p_1$ & $p_2$ & $\gamma^*$ & $\gamma_{min}$ & B       & r             &$\delta$   &      disk Luminosity  &               $n_e$  \\
       &       &       &            &               & (Gauss) & ($10^{16}$ cm) &
       & ($10^{45} erg\ cm^{-2}\ s^{-1}$) &          ($e^-/cm^3$) \\
\hline
first  & 2     &  5    & 200        &  3            & 12      &  2            & 9        &  6                   & $150$               \\
second & 2     &  4.7  & 300        &  3            & 10      &  2            & 9        &  6                   & $\ 70$               \\
\hline
\end{tabular}
\end{table*}

In the scenario proposed by \cite{sikora2001}, during the acceleration phase,
the accelerated electrons population increases, saturating at high energy first.
When the  phase of electrons acceleration stops, the energy break $\gamma ^*$ of
the electrons population moves to lower energies, 
reaching the critical energy $\gamma _C$ (balancing the radiative
cooling time with the duration of the acceleration period) or even lower
values. In that model, the gamma-ray light curve
reaches its maximum before the hard X, then decay faster than hard X-ray light curve. 
That scenario might be able to fit the data of our multiwavelength campaign,
provided that the second week is related to an electrons acceleration phase,
and the first week to the late phase of a previous episode. Thence the gamma-ray activity
and the high value of $\gamma^*$, during the second week of observation
are the signature of the acceleration phase.

%

\section{Summary and Conclusions}


We presented data of a pre-scheduled 3-week multi-wavelength
campaign on 3C 273 carried out between mid-December 2007 and
January 2008, covering from the near-Infrared to the gamma-ray
energy bands, for the first time after the demise of the EGRET
instrument. The source was found in high state in the X-rays, with
a 5-100 keV flux a factor of $\sim$3 higher than the typical value
in historical observations (e.g., \cite{corvo} for the INTEGRAL
data). Instead, the AGILE gamma-ray data showed a flux lower-equal
to the EGRET measurements, and the optical/IR measurements
provided fluxes very similar to the "standard values"
for this source.\\

Our multi-frequency and continuous set of data allowed us to study
the short-term variability (days to week) of this source. The
simultaneous light curves from the different instruments do not
show any strong correlation, except for an indication of an
anti-correlated variability between X-rays and gamma-rays: all the
soft and hard X-ray measurements show a decreasing trend at the
time of our single positive detection in the gamma-rays in the
second week of observation, preceded and followed by
non-detections in the first and third week of our campaign.\\

This behavior can be interpreted and understood when we use our
multi-frequency data to model the source spectral energy
distribution. Using a model composed by a one-zone homogeneous
Synchrotron-Self-Compton plus external Compton from an accretion
disk, we find that the spectral variability between the first and
the second week is consistent with an acceleration episode of the
electron population responsible for the synchrotron emission.
In our model the detectable synchrotron variations are
in the far-IR and in the soft X-rays, where we didn't have adequate coverage,
whereas the near-IR and optical remain almost unchanged.
But the signature of the acceleration is brought
up by the inverse Compton peak in the X and gamma-ray energy
ranges.

We note that shifts of the inverse-Compton peak from observation
to observation were previously proposed (see \cite{mcnaron}) from
the comparison of the June 1991 multi-wavelength campaign, and the
OSSE observation of September 1994. Our multi-frequency
observation and modelling suggests that this behaviour is a more
general feature of this source, happening on shorter timescales.

Our observation of a weaker X-ray flux in the second week
motivated us to study the Seyfert-like disk reflection hump in
this source. The wide band spectral data from the INTEGRAL
instruments show that the jet (non-thermal) emission alone does
not describe the energy spectrum adequately. A reflection hump
improves the X-ray spectral modelling. We then found that in the
second week the jet contribution to the X-ray emission gets
dimmer, due to the shift to higher energy of the electron
population discussed above, making the likely constant disk
contribution to emerge. The quality of our data did not allow to
put any constraints on the possible variability of the reflection
component, that in our data is consistent with an intermediate
intensity reported form previous observations (e.g.,
\cite{grandi2004})

\begin{acknowledgements}
The AGILE mission is funded by the Italian Space Agency (ASI) with scientific
and programmatic participation by the Italian Institute of Astrophysics (INAF)
and the Italian Institute of Nuclear Physics (INFN).\\ 
We thank the Swift PI N. Gehrels for the approval of the ToO observations, and
the Swift team for performing them. We are indeed very grateful to the
INTEGRAL Science Operation Center (ISOC) and INTEGRAL Science Data Center (ISDC) teams for the optimal
scheduling of the INTEGRAL pointings, and the support and the prompt alerts
during the campaign. We thank A. Bazzano for her help in the organization of
the campaign, and A. De Rosa for her suggestions.
\end{acknowledgements}

\begin{appendix}
\newpage
\section{Complementary data}

\longtab{1}{
\begin{longtable}{c c c c c }
\caption{\label{tab:hardx_lc_app} Hard X data from SuperAGILE and ISGRI. The Flux is reported in mCrab units in the energy band 20 - 60 keV}\\
\hline\hline
start date & stop date & exposure &flux    & observatory \\
(MJD)      & (MJD)     & (ks)     &(mCrab) &             \\ \hline
\endfirsthead
\caption{continued.}\\
\hline\hline
start date & stop date & exposure&flux    & observatory \\
(MJD)      & (MJD)     & (ks)    &(mCrab) &             \\ \hline
\endhead
\hline
\endfoot

54450.72 & 54453.91 & 136 & 30.2 $\pm$ 2.9 & SuperAGILE \\
54453.91 & 54457.10 & 141 & 24.5 $\pm$ 3.0 & SuperAGILE \\
54458.30 & 54461.63 & 140 & 21.6 $\pm$ 2.2 & SuperAGILE \\
54461.63 & 54464.96 & 150 & 21.2 $\pm$ 2.4 & SuperAGILE \\
54469.57 & 54473.46 & 176 & 22.2 $\pm$ 2.3 & SuperAGILE \\ \hline

54453.86 & 54454.14 & 21 & 25.7 $\pm$ 1.6 & ISGRI \\
54454.14 & 54454.43 & 18 & 22.4 $\pm$ 1.6 & ISGRI \\
54454.43 & 54454.72 & 21 & 24.2 $\pm$ 2.4 & ISGRI \\
54454.72 & 54455.01 & 22 & 22.4 $\pm$ 1.5 & ISGRI \\
54455.01 & 54455.30 & 22 & 23.1 $\pm$ 1.5 & ISGRI \\
54455.30 & 54455.59 & 24 & 23.8 $\pm$ 1.4 & ISGRI \\
54455.59 & 54455.88 & 22 & 25.1 $\pm$ 1.5 & ISGRI \\
54455.88 & 54456.17 & 22 & 24.0 $\pm$ 1.5 & ISGRI \\
54456.17 & 54456.38 & 14 & 26.0 $\pm$ 2.9 & ISGRI \\

54459.73 & 54459.93 &  8 & 18.3 $\pm$ 2.6 & ISGRI \\
54459.93 & 54460.22 & 24 & 21.1 $\pm$ 1.5 & ISGRI \\
54460.22 & 54460.51 & 23 & 19.3 $\pm$ 1.6 & ISGRI \\
54460.51 & 54460.80 & 24 & 18.2 $\pm$ 1.5 & ISGRI \\
54460.80 & 54461.09 & 20 & 17.9 $\pm$ 1.6 & ISGRI \\
54461.09 & 54461.38 & 23 & 20.1 $\pm$ 1.6 & ISGRI \\
54461.38 & 54461.67 & 24 & 22.2 $\pm$ 1.5 & ISGRI \\
54461.67 & 54461.96 & 15 & 20.3 $\pm$ 1.9 & ISGRI \\
54461.96 & 54462.25 & 24 & 20.2 $\pm$ 1.5 & ISGRI \\
54462.25 & 54462.27 &  6 & 17.6 $\pm$ 4.1 & ISGRI \\

54465.72 & 54466.01 & 16 & 23.2 $\pm$ 1.9 & ISGRI \\
54466.01 & 54466.30 & 24 & 20.4 $\pm$ 1.4 & ISGRI \\
54466.30 & 54466.59 & 23 & 19.1 $\pm$ 1.5 & ISGRI \\
54466.59 & 54466.88 & 24 & 24.9 $\pm$ 1.5 & ISGRI \\
54466.88 & 54467.17 & 23 & 21.7 $\pm$ 1.5 & ISGRI \\
54467.17 & 54467.45 & 24 & 27.1 $\pm$ 1.5 & ISGRI \\
54467.45 & 54467.74 & 24 & 22.6 $\pm$ 1.5 & ISGRI \\
54467.74 & 54468.03 & 22 & 23.0 $\pm$ 1.5 & ISGRI \\
54468.03 & 54468.18 & 21 & 26.1 $\pm$ 1.9 & ISGRI \\
\end{longtable}
}

\longtab{2}{
\begin{longtable}{c c c c c c c c}
\caption{\label{tabapp:rem_lc_app} REM and UVOT data for the multiwavelength campaign}\\
\hline\hline
date       &  MJD     & filter & $\lambda$     & exposure & magn & energy flux & observatory \\
(aaaammdd) &          &        & ($\AA$)       &  (s)     &      & (mJy)       &             \\ \hline
\endfirsthead
\caption{continued.}\\
\hline\hline
date       &  MJD     & filter & $\lambda$     & exposure & magn & energy flux & observatory \\
(aaaammdd) &          &        & ($\AA$)       &  (s)     &      & (mJy)       &             \\ \hline
\endhead
\hline
\endfoot

20080106 & 54471.5 & UVM2 & 2231  & 729 & 11.16 $\pm$ 0.03  & 31.0 $\pm$  0.9 & UVOT \\
20080104 & 54469.7 & UVM2 & 2231  & 610 & 11.17 $\pm$ 0.03  & 30.7 $\pm$  0.9 & UVOT \\ \hline
20080106 & 54471.5 & B    & 4329  & 268 & 12.86 $\pm$ 0.02  & 33.2 $\pm$  0.6 & UVOT \\ \hline
20080106 & 54471.5 & V    & 5402  & 268 & 12.67 $\pm$ 0.01  & 33.1 $\pm$  0.3 & UVOT \\
20080104 & 54469.7 & V    & 5402  & 213 & 12.63 $\pm$ 0.01  & 34.4 $\pm$  0.3 & UVOT \\ \hline

20080114 & 54479.3 & V &  5496 & 300 & 12.64 $\pm$ 0.02 &  33.9 $\pm$  0.7 & REM \\
20080111 & 54476.3 & V &  5496 & 300 & 12.65 $\pm$ 0.02 &  33.7 $\pm$  0.6 & REM \\
20080106 & 54471.3 & V &  5496 & 300 & 12.53 $\pm$ 0.03 &  37.5 $\pm$  0.9 & REM \\
20080102 & 54467.4 & V &  5496 & 300 & 12.61 $\pm$ 0.04 &  35.1 $\pm$  1.2 & REM \\
20071227 & 54461.3 & V &  5496 & 300 & 12.70 $\pm$ 0.06 &  32.2 $\pm$  1.7 & REM \\
20071223 & 54457.3 & V &  5496 & 300 & 12.63 $\pm$ 0.06 &  34.4 $\pm$  1.8 & REM \\
20071220 & 54454.3 & V &  5496 & 300 & 12.58 $\pm$ 0.02 &  35.9 $\pm$  0.7 & REM \\
20071211 & 54445.3 & V &  5496 & 300 & 12.67 $\pm$ 0.06 &  33.3 $\pm$  1.9 & REM \\ \hline

20080114 & 54479.3 & I &  7895 & 300 & 12.04 $\pm$ 0.05 &  40.3 $\pm$  1.8 & REM \\
20080111 & 54476.3 & I &  7895 & 300 & 12.10 $\pm$ 0.05 &  38.1 $\pm$  1.9 & REM \\
20080106 & 54471.3 & I &  7895 & 300 & 12.02 $\pm$ 0.03 &  41.1 $\pm$  1.0 & REM \\
20080102 & 54467.4 & I &  7895 & 300 & 12.01 $\pm$ 0.03 &  41.5 $\pm$  1.2 & REM \\
20071227 & 54461.3 & I &  7895 & 300 & 12.14 $\pm$ 0.04 &  37.0 $\pm$  1.3 & REM \\
20071223 & 54457.3 & I &  7895 & 300 & 12.14 $\pm$ 0.03 &  37.0 $\pm$  1.0 & REM \\
20071220 & 54454.3 & I &  7895 & 300 & 12.06 $\pm$ 0.02 &  39.8 $\pm$  0.8 & REM \\
20071211 & 54445.4 & I &  7895 & 300 & 12.09 $\pm$ 0.03 &  38.7 $\pm$  1.2 & REM \\ \hline

20080114 & 54479.3 & R &  6396 & 300 & 12.48 $\pm$ 0.02 &  33.0 $\pm$  0.7 & REM \\
20080111 & 54476.3 & R &  6396 & 300 & 12.52 $\pm$ 0.01 &  31.8 $\pm$  0.3 & REM \\
20080106 & 54471.3 & R &  6396 & 300 & 12.40 $\pm$ 0.03 &  35.5 $\pm$  0.8 & REM \\
20080102 & 54467.4 & R &  6396 & 300 & 12.49 $\pm$ 0.04 &  32.7 $\pm$  1.1 & REM \\
20071227 & 54461.3 & R &  6396 & 300 & 12.55 $\pm$ 0.05 &  31.1 $\pm$  1.5 & REM \\
20071223 & 54457.3 & R &  6396 & 300 & 12.47 $\pm$ 0.06 &  33.3 $\pm$  1.9 & REM \\
20071220 & 54454.3 & R &  6396 & 300 & 12.45 $\pm$ 0.02 &  34.0 $\pm$  0.7 & REM \\
20071211 & 54445.4 & R &  6396 & 300 & 12.49 $\pm$ 0.03 &  32.8 $\pm$  0.9 & REM \\ \hline

20080114 & 54479.3 & J & 12596 &  30 & 11.53 $\pm$ 0.05 &  39.8 $\pm$  1.9 & REM \\
20080111 & 54476.3 & J & 12596 &  30 & 11.45 $\pm$ 0.06 &  42.8 $\pm$  2.2 & REM \\
20080106 & 54471.3 & J & 12596 &  30 & 11.22 $\pm$ 0.08 &  53.0 $\pm$  3.7 & REM \\
20080102 & 54467.4 & J & 12596 &  30 & 11.47 $\pm$ 0.03 &  42.1 $\pm$  1.4 & REM \\
20071230 & 54464.3 & J & 12596 &  30 & 11.47 $\pm$ 0.04 &  42.0 $\pm$  1.4 & REM \\
20071227 & 54461.3 & J & 12596 &  30 & 11.53 $\pm$ 0.03 &  39.9 $\pm$  1.3 & REM \\
20071223 & 54457.3 & J & 12596 &  30 & 11.55 $\pm$ 0.03 &  39.1 $\pm$  1.2 & REM \\
20071220 & 54454.3 & J & 12596 &  30 & 11.47 $\pm$ 0.04 &  41.9 $\pm$  1.4 & REM \\
20071211 & 54445.3 & J & 12596 &  30 & 11.42 $\pm$ 0.05 &  44.1 $\pm$  2.1 & REM \\ \hline

20080114 & 54479.3 & H & 15988 &  30 & 10.71 $\pm$ 0.04 &  56.6 $\pm$  2.1 & REM \\
20080111 & 54476.3 & H & 15988 &  30 & 10.68 $\pm$ 0.05 &  58.4 $\pm$  2.4 & REM \\
20080106 & 54471.3 & H & 15988 &  30 & 10.54 $\pm$ 0.07 &  66.7 $\pm$  4.5 & REM \\
20080102 & 54467.4 & H & 15988 &  30 & 10.65 $\pm$ 0.02 &  59.8 $\pm$  1.3 & REM \\
20071230 & 54464.3 & H & 15988 &  30 & 10.68 $\pm$ 0.02 &  58.5 $\pm$  1.2 & REM \\
20071227 & 54461.3 & H & 15988 &  30 & 10.69 $\pm$ 0.02 &  58.0 $\pm$  1.3 & REM \\
20071223 & 54457.3 & H & 15988 &  30 & 10.71 $\pm$ 0.02 &  56.6 $\pm$  1.2 & REM \\
20071220 & 54454.3 & H & 15988 &  30 & 10.65 $\pm$ 0.03 &  59.8 $\pm$  1.5 & REM \\
20071211 & 54445.4 & H & 15988 &  30 & 10.70 $\pm$ 0.04 &  57.5 $\pm$  2.1 & REM \\ \hline

20080114 & 54479.3 & K & 22190 &  30 &  9.70 $\pm$ 0.05 &  88.8 $\pm$  4.2 & REM \\
20080111 & 54476.3 & K & 22190 &  30 &  9.55 $\pm$ 0.08 & 102.0 $\pm$  7.5 & REM \\
20080102 & 54467.4 & K & 22190 &  30 &  9.65 $\pm$ 0.03 &  93.2 $\pm$  2.3 & REM \\
20071230 & 54464.4 & K & 22190 &  30 &  9.68 $\pm$ 0.03 &  91.0 $\pm$  2.2 & REM \\
20071227 & 54461.3 & K & 22190 &  30 &  9.67 $\pm$ 0.03 &  91.1 $\pm$  2.3 & REM \\
20071223 & 54457.3 & K & 22190 &  30 &  9.56 $\pm$ 0.03 & 101.3 $\pm$  2.9 & REM \\
20071220 & 54454.3 & K & 22190 &  30 &  9.58 $\pm$ 0.11 &  99.6 $\pm$ 10.1 & REM \\
20071211 & 54445.4 & K & 22190 &  30 &  9.51 $\pm$ 0.05 & 105.9 $\pm$  4.8 & REM \\

\end{longtable}
}

\end{appendix}

\end{document}